# Comparison of the impacts of three types of plasma-activated water on the seed germination and plant growth of lettuce (*Lactuca sativa*).

Ramin Mehrabifard [1], Adriana Mišúthová[1], Zdenko Machala[1*]

[1]Division of Environmental Physics, Faculty of Mathematics, Physics and Informatics, Comenius University, Mlynská dolina, 842 48 Bratislava, Slovakia

[*]Correspondence: machala@fmph.uniba.sk

**Abstract**

Cold air plasma typically generates reactive oxygen and nitrogen species (RONS), which are transported into water to produce plasma-activated water (PAW). This work examines the effect of PAW produced by three different plasma systems on lettuce: transient spark, fountain dielectric barrier discharge, and microwave plasma. Physiochemical PAW properties and concentrations of RONS (ozone, hydrogen peroxide, nitrite, and nitrate) were measured. Seed germination parameters were recorded in 8-day paper tests. The effect of PAW irrigation was investigated after six weeks of plant growth in the soil by measuring the germination rate, plant and root length, dry and fresh plant weight, number of leaves, and photosynthetic pigments. PAW, dependent upon its RONS contents, enhances plant development and affects its physiological parameters.

1. Introduction

In many parts of the world, agricultural cultivation is a significant economic sector. Fresh, high-quality agricultural goods with an appropriate amount of macro- and micronutrients are in high demand because they provide a decent quality of life. Many inorganic and organic fertilizers, which supply plant nutrients required for their development, are used to promote the growth of crops, fruits, and vegetables [1]. In addition to fertilizers, several physical techniques have also been used to promote the agricultural plant growth, such as pulsed electric fields or static magnetic fields [2], [3]. Cold (nonthermal, non-equilibrium) plasma (CP) also represents an innovative physical approach for plant growth promotion.

CP formed by different types of atmospheric pressure air discharges generates a very reactive atmosphere consisting of energetic electrons, radicals, and numerous gaseous reactive species such as •OH, $H_2O_2$, $NO_x$, and $HNO_x$ [4], [5]. Upon interaction with water, these species dissolve into the water and generate plasma-activated water (PAW). Long-lived reactive oxygen and nitrogen species (RONS) like $H_2O_2$, $NO_2^-$, or $NO_3^-$ are the most typical constituents of PAW, which can be regarded as a sustainable and clean substitute for chemical fertilizers. CP and PAW have a possibility to be efficiently employed in numerous agricultural uses [6]–[8].

Most of the current research of plasma agriculture application focuses on seed disinfection and enhanced germination, as well as promotion of the plant development, and other applications involving direct plasma treatment [9]–[13]. There is a growing interest in indirect plasma treatments on plants and seedlings, especially by the application of PAW, or plasma activated fertilizer, [14]–[19].

$H_2O_2$, $NO_2^-$, and $NO_3^-$ are the long-lived plasma-generated RONS typically present in the PAW. They may be readily incorporated into plant cells where they can function as signaling molecules in plant physiology, or as sources of essential nitrogen nutrients for the plant. $H_2O_2$ may pass through a cell membrane either by free diffusion or through membrane proteins called aquaporins that help move water through the membrane [20]. $H_2O_2$ may be scavenged in the cell by the enzyme catalase, which produces oxygen and water, or by the antioxidant enzyme peroxidase. Since hydrogen peroxide is the most persistent reactive oxygen species (ROS), it is essential for intracellular communication in a variety of physiological functions.

One important source of nitrogen, a macronutrient that is vital to plants, is nitrate ($NO_3^-$). It is transported into plant cells by specific membrane transport proteins or concentration-dependent diffusion, which is a slower process [21], [22]. Furthermore, $NO_3^-$ also functions as a signaling molecule, triggering the activation of genes necessary for absorption and self-transport inside the plant cell. Particularly for active photosynthesis and absorption in support of the biomass formation, the absorbed nitrogen might be used effectively. Certain plants can also use $NO_2^-$ (nitrite) as a source of nitrogen; however, it may also be hazardous to many plants [23].

According to Gierczik et al. [24], the presence of $H_2O_2$ and $NO_3^-$ in PAW improved seedling resistance to simultaneous low-temperature and hypoxic stress conditions during germination. Zhang et al. [25] demonstrated that they achieved 80% germination rates rather than 30% by employing plasma-activated tap water. Additionally, when compared to industrial synthetic fertilizer, plasma-activated tap water produced better rates of stem elongation and ultimate stem lengths. They concluded that these gains were mostly dependent on the mixture of two particular long-lived PAW species. Fan et al. [26] used deionized water exposed to CP in order to examine the impact of PAW on mung bean sprouts. By assessing factors including germination rate, growth characteristics, total phenolic and flavonoid levels, and antioxidant enzyme activity, they assessed the PAW impact and concluded that it might promote mung seed germination and growth.

In this work we use three different plasma sources to generate RONS in water with different concentrations to investigate the effect of each species ($NO_3^-$, $NO_2^-$, $H_2O_2$) on lettuce seed and plant parameters separately. Prior to using PAW for the irrigation of target plant seedlings, concentrations of long-lived species ($O_3$, $H_2O_2$, $NO_2^-$, $NO_3^-$) were analyzed by UV–VIS absorption spectroscopy, along with the physiochemical parameters of PAW, such as pH, Oxidation-Reduction Potential (ORP), Total Dissolved Solids (TDS), Electrical Conductivity (EC), and temperature. Although all three types of PAW contain a mixture of RONS, water activated by MW plasma generates high concentrations of $NO_3^-$, Transient spark (TS) generates the highest $NO_2^-$, and fountain dielectric barrier discharge (FDBD) provides lower RONS concentration than TS. $O_3$ concentration for all these PAW types is small. We examined specific germination and plant growth characteristics (germination, plant and root length, number of leaves, as well as fresh and dry weight) and physiological data, including the concentration of photosynthetic pigments, to assess the efficacy of PAW in enhancing germination and plant growth.

## 2. Materials and methods

### 2.1. Plasma setups

Experimental setups are shown in figures 1, 2, and 3. Three different discharge reactors are used for this experiment: Multi-pin transient spark (TS) discharge, fountain dielectric barrier discharge (FDBD), and microwave (MW) plasma. A high voltage probe (Tektronix P6015A) is used for measuring voltage. A Rogowski current meter (Pearson Electronics 2877) measures the discharge current. Digitizing oscilloscope Tektronix TDS 2024 records and processes the time-dependent waveform of the electrical characteristics of the discharge (voltage and current). The same batch of tap water with a typical initial conductivity of ~500 µS/cm was used to make all the three types of PAW treatments, which also served as the untreated control (TW). This makes sure that plasma activation is the only process that changes the properties of the water. In contrast to deionized or distilled water, it is more abundant, more physiologically suitable for plants, and has a natural buffering capacity to maintain a stable pH due to its bicarbonate buffer system. PAWs were used for irrigation within 10 minutes of plasma activation to preserve their reactive species content.

#### 2.1.1 Multi-pin transient spark (TS)

Single TS discharge has been extensively investigated in our research group, including the setups with liquid electrode [27]–[29]. Here we employ a Multi-pin TS that operates in pin-to-plane geometry with a common liquid grounded electrode. The power needle electrodes are connected to a DC power supply each separately through a 4.7 *MΩ* ballast resistor and 50 *pF* external capacitor. The discharge gap, from the tip of needle electrodes to the water surface is ~1 *cm* long. The plasma activation duration for water was established at 1 mL min $^{-1}$ per electrode, meaning each milliliter of water was subjected to plasma activation for 1 minute, which was kept constant throughout all tests. The batch volume of 210 mL of water was treated by TS plasma for 10 min by 21 electrodes. Figure 1 a and b shows a schematic diagram and a real picture of the multi-pin TS plasma source, and Figure 1 c and d show voltage and current-voltage waveform of the discharge, respectively.

The TS plasma power was measured using voltage, current, and frequency. The instantaneous power, P(t), at any time t is given by:

$$P(t) = V(t).I(t)$$

Where *V(t)* is the instantaneous voltage, and *I(t)* is the instantaneous current. To find the average power over one period, we integrate *P(t)* over the period (T) and divide by the T, i.e. multiply by the frequency (f) :

$$P_{avg} = \frac{1}{T}\int_0^T P(t)dt = f\int_0^T V(t).I(t)\,dt$$

The typical average power of the Multipin TS is 52.5 W, i.e. 2.5 W per needle.

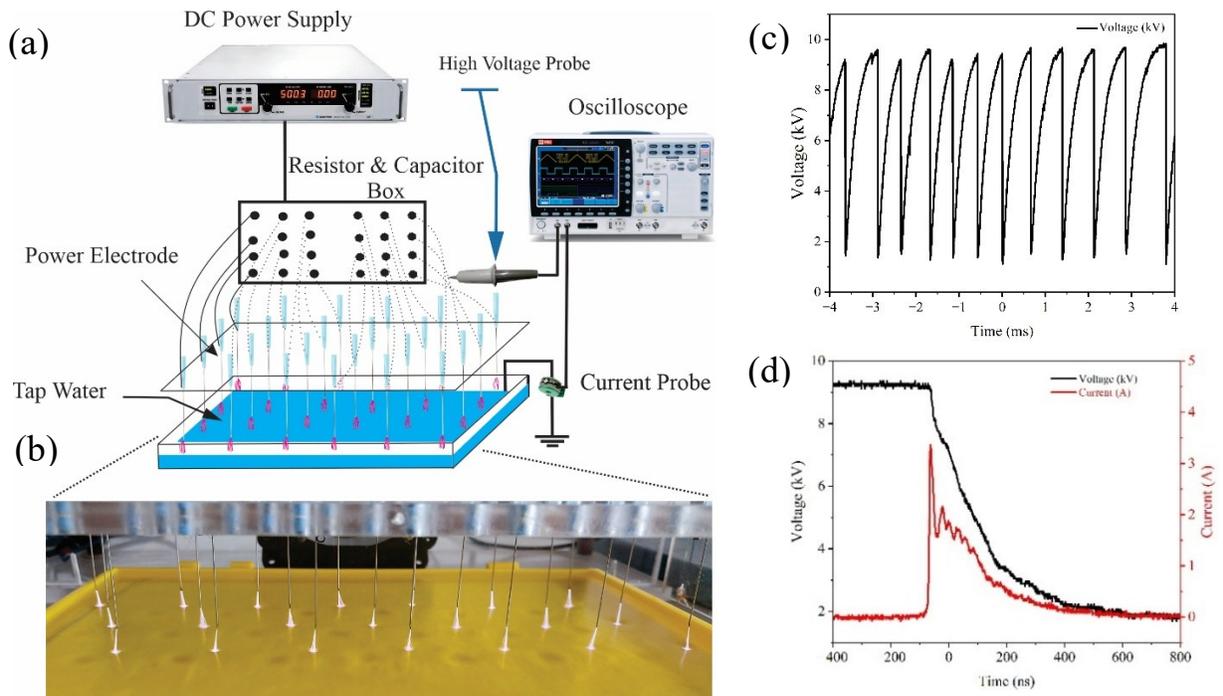

Figure 1. Multi-Pin TS discharge, (a) schematic diagram, (b) real photograph, (c) voltage waveform on ms timescale, and (d) current-voltage waveform on 100 ns timescale.

### 2.1.2 Fountain dielectric barrier discharge (FDBD)

Figure 2 a and b show real and schematic diagram of FDBD reactor and water circulating system, respectively. The same reactor has been used previously in [30]. An AC neon transformer (15 kV, 20 kHz) and a variable transformer are used for generating plasma. A DC power supply was used for the water pump. The reactor is suitable to treat a large volume of water with a typical 85 *mL/min* flow rate. The reactor is shaped like a coaxial cylinder, and the dielectric is a quartz tube that is 180 × 25 *mm* in size with 1.5 *mm* thickness. The central tube electrode is made of copper, and at the outer surface, a copper coil is wrapped around the glass dielectric. The spacing between the inner and outer tubes (discharge gap) is 3 *mm*. After being fed into the reactor, water flows laminarly upwards through the central Cu tube before dropping out the top of the central tube into the gap with a micro discharge zone. One liter of water was treated for 20 minutes. Figure 2 c shows the current-voltage waveform of the FDBD. We used the Lissajous figure method to measure the dissipated power. To measure the charge that was transferred during plasma discharge, a reference capacitor ($C_m$ = 6 nF) was placed in series with the FDBD device. The voltage across the FDBD electrodes and the voltage across the measurement capacitor were simultaneously recorded using the high voltage probes and a digital oscilloscope. The charge $Q$ was calculated from the voltage across the capacitor using the relation $Q = C_m \cdot V_C$. A Lissajous figure was plotted by DBD voltage (x-axis) against the calculated charge (y-axis) over one full AC cycle. The area enclosed by the resulting figure corresponds to the energy dissipated per cycle (figure 2 d), which corresponds to the discharge power of 48.8 W.

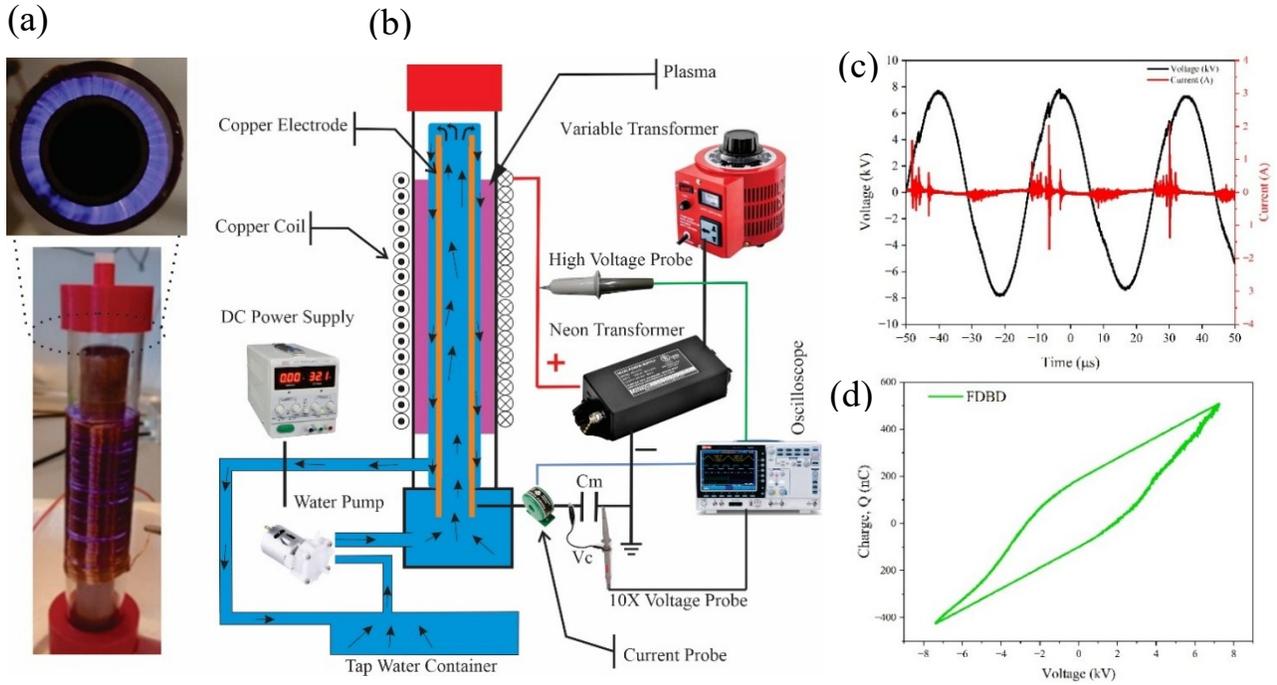

Figure 2. Fountain Dielectric Barrier Discharge, (a) real photograph, (b) schematic diagram, (c) voltage and current waveforms, (d) Lissajous figure.

### 2.1.3. Microwave (MW) plasma

The water activated by MW plasma is prepared by *Proline solutions* company (Austria). The microwave power supply operates at a frequency of 2.45 *GHz*. Schematic diagram of the MW plasma is shown in Figure 3. The plasma system consumes 1.8 kW of electrical power and produces 45 L of PAW per hour. More details of the MW plasma activated water are specified in reference [31].

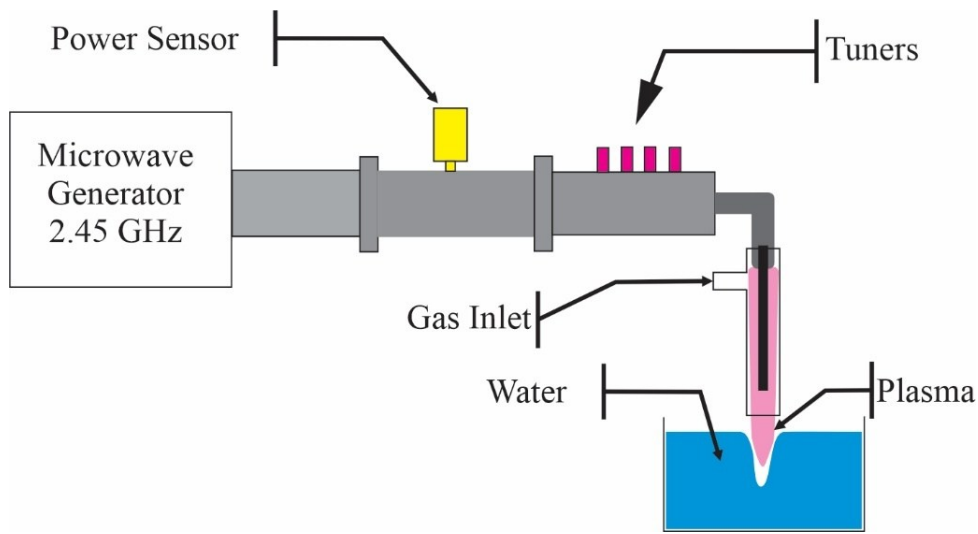

Figure 3. Schematic of Microwave plasma set up in *Proline solutions* company (Austria)

## 2.2. Plasma activated water analysis

### 2.2.1. Physiochemical parameters of PAW

The pH and ORP are measured using a portable meter (WTW 3110, Weilheim, Germany). EC and temperature are measured with a digital conductivity meter (GMH 3430, GREISINGER electronic, Germany). TDS is measured by 4-in-1 Multi Meter (Noyafa Digital, Philippines).

### 2.2.2. Hydrogen peroxide and ozone measurements

The quantity of $H_2O_2$ in PAW was measured by its interaction with titanyl ions from titanium oxysulfate ($TiOSO_4$), yielding a yellow product of pertitanic acid with a maximum absorbance peak at 407 nm [32]:

$$Ti^{4+} + H_2O_2 + 2H_2O \rightarrow TiO_2H_2O_2 + 4H^+ \qquad (1)$$

UV-VIS spectrometer (Shimadzu 1900) is used for detecting the absorption spectra. The yellow-colored compound generated is stable for a minimum of 6 hours. To counteract the decomposition of $H_2O_2$ by the reaction with nitrites in PAW, a 60-mM solution of sodium azide ($NaN_3$) is included into the samples containing $H_2O_2$ prior to their mixing with $TiOSO_4$ reagent. Sodium azide promptly converts nitrites into molecular nitrogen under acidic conditions.

$$3N_3^- + NO_2^- + 4H^+ \rightarrow 5N_2 + 2H_2O \qquad (2)$$

After the combination of the samples with azide, we add the titanium oxysulfate reagent into the sample. The $TiOSO_4$ to the sample ratio is 2:1.

We used the indigo colorimetric technique using indigo blue reagent to detect the quantity of dissolved ozone in PAW [33]. This method relies on the fast and precise interaction between ozone and indigo blue dye in an acidic environment, leading to the degradation of the indigo blue molecular structure, which causes the dye to fade or bleach. The bleaching action results in a quantifiable reduction of the peak absorbance of indigo blue at 600 nm. The extent of the absorbance reduction is directly proportional to the concentration of ozone in the sample. Consequently, by quantifying the variation in absorbance using a UV-Vis spectrophotometer (Shimadzu 1900), the quantity of dissolved ozone can be precisely determined.

### 2.2.3. Nitrite and nitrate measurements

Nitrite $NO_2^-$ was quantified using a commercial kit containing Griess reagent (Cayman Chemicals, Ann Arbour, MI, USA), resulting in a pink azo-product with a maximal absorbance peak at 540 nm [34]. The $NO_2^-$ concentration in PAW is proportional to this absorbance and is calculated from the calibration curve.

For nitrate $NO_3^-$ measurement in PAW we used 2,6- dimethylphenol (DMP) (Sigma-Aldrich, USA) that with nitrate in sulfuric and phosphoric acid solution reacts to form 4-nitro-2,6-dimethylphenol that is spectrophotometrically detected at ~330 nm [35]. The $NO_3^-$ concentration in PAW is proportional to this absorbance and is calculated from the calibration curve. However, when PAW contains both $NO_2^-$ and $NO_3^-$, the DMP chemical probe is also sensitive to $NO_2^-$, hence careful measurement of $NO_2^-$ by the Griess reagent and its subtraction from the DMP absorbance was applied.

## 2.3. Effect of PAW on seeds and plants

### 2.3.1. Seed germination parameters

*Lactuca sativa* (hybrid Cassini) seeds were purchased from regional agriculture market (Moravoseed, Slovakia). The germination experiments were performed in petri dishes using filter paper (KA 4, Czech Republic), we used two filter papers in each Petri dish. The seeds were divided into four experimental groups: tap water as the control, and plasma-activated water produced by TS, FDBD and MW plasma. The samples were irrigated with 3 mL of PAW at 24-hour intervals and incubated for 7 days. Each group consisted of three replicates (30 seeds each), and all three replicated experiments were conducted under identical experimental conditions to verify the repeatability of the results. The seeds were germinated in a controlled environment at a temperature of 23±2 °C, for 24 hours in darkness. Petri dishes were briefly exposed to ambient room light solely for counting of the germinated seeds (less than 1 min). The number of germinated seeds was monitored and documented daily.

- Germination percentage (GP)
  GP is typically expressed as the percentage of seeds that successfully germinate out of the total number of planted seeds, over a specific period of time:
  $$GP = \frac{N_g}{N_t} \times 100 \quad (3)$$
  $N_g$ is the quantity of germinated seeds, $N_t$ is total number of seeds.
- Mean Germination Rate (MGR)

MGR is a measure that takes into account not just how many seeds germinate, but also the time it takes for the seeds to germinate. It provides a more comprehensive understanding of the germination process.

$$MGT = \frac{\sum(n \times t)}{\sum n} \quad (4)$$

Where *n* is number of seeds germinated at time t.

- Seedling Vigor Index (SVI)

The Seedling Vigor Index of seedlings is a metric used to assess the general health, vigor, and potential for effective development of plant seedlings. It combines two essential elements:

$$SVI = \text{Germination percentage} \times \text{shoot length} \quad (5)$$

- Richard function

The effect of stressors on germination can be estimated by induced changes in parameters of germination kinetics, derived using Richard's function [36] applied for the analysis of the germination seed population [37]. This function was chosen over simpler models (e.g., logistic or Gompertz) due to its ability to represent asymmetric germination curves:

$$Y_t = \frac{a}{(1 + b.d.e^{-c.t})^{1/d}} \quad (6)$$

where $Y_t$ is gemination parameters; *a*, *b*, *c*, and *d* are fitting parameters; t is time. The parameter in the equation is determined based on the percentage of fully germinated seeds in the sample. The optimal value of parameters *a, b, c,* and d selected by applying the nonlinear least squares method and the "Solver" in Microsoft Excel and goodness of fit ($R^2$) values were recorded for each treatment group. The indices of germination kinetics include the final germination percentage ($V_i$), median germination ($M_e$) and quartile deviation ($Qu$):

$$V_i = a \quad (7)$$

$$M_e = \frac{1}{c}.\ln\frac{b.d}{2^d - 1} \quad (8)$$

$$Qu = \frac{1}{2c}.\ln\frac{4^d - 1}{(4/3)^d - 1} \quad (9)$$

$$SK = 2.\left\{\ln\frac{d}{2^d - 1} / \ln\frac{4^d - 1}{(4/3)^d - 1}\right\} \quad (10)$$

The $M_e$ is inversely related to the germination rate, meaning that lower $M_e$ value corresponds to a higher germination rate, as it signifies that half of the seeds have germinated in less time. The $Qu$ indicates the dispersion of germination time in a seed lot. SK denotes the skewness in the frequency distribution of germination time. These parameters are calculated for control and treat groups using the coefficients *a, b, c,* and d of Richards functions.

## 2.3.2. Physical parameters of plant growth experiments in soil

The same *Lactuca sativa* (hybrid Cassini) seeds from regional agriculture market (Moravoseed, Slovakia) were used for the experiments where the seeds were planted into pots with a soil and allowed to grow. Over the course of 6 weeks, they were irrigated with PAW every two to three days. 5 ml of PAW is used for irrigation before germination, 20 ml during the second and third weeks, and 30 ml during the fourth week. There were 8 seeds in each pot and 12 pots for each group. Each pot (8×8 cm) included 8 seeds spaced evenly to minimize competition. Although the density of seeds per pot is relatively high, the same setup was used across all treatment groups. This procedure was enough for short-term (6-week) early growth analyses, though it may not reflect optimal spacing for full development. This limitation is acknowledged when interpreting biomass-related results. Following six weeks of irrigation with the three types of PAW a comprehensive assessment of visual characteristics (quantity and quality of leaves), plant height, plant growth metrics (fresh and dry weight of whole plant), germination rate in soil, and photosynthetic pigments (chlorophyll a + b) content in both the above-ground parts and roots of lettuce plants were conducted. In this experiment the pots with plants were located at a west-oriented window at room temperature (20 °C) and approximately 16/8 daylight/night cycle (with ~7 h of direct sunlight daily in average).

### 2.3.3. Photosynthetic pigment concentration measurement

Lichtenthaler's method [38] was used to determine the concentration of photosynthetic pigments in leaves, namely chlorophylls a and b and carotenoids. The samples of lettuce leaves tissue are homogenized with 2ml of 80% ethanol and centrifuged at 14000 g for 15 min. The supernatant was adjusted to a certain amount and diluted to achieve a linear absorbance range of 0.3–0.7. The concentrations of chlorophyll a, chlorophyll b, and carotenoids (xanthophylls and carotenes, x + c) were assessed based on absorbance readings obtained from a UV–VIS spectrophotometer (Shimadzu 1900, Japan) at wavelengths of 664, 648, and 470 nm, respectively. Three samples from each treatment were collected in every experimental trial.

### 2.4. Statistics

The results are presented using mean values ± standard deviation. A one-way analysis of variance (ANOVA) and subsequent multiple range test using the least significant difference technique (LSD) were conducted to assess the differences across the groups. The various lowercase letters indicate a statistically significant difference at a probability level of $p < 0.05$.

### 3. Results and discussion

### 3.1. PAW characterization

PAW serves as a reservoir of various RONS, which have been shown to enhance the plant growth under stress conditions while also offering a potential alternative to conventional fertilizers. The priming effect of hydrogen peroxide ($H_2O_2$) on seeds plays a crucial role in accelerating and amplifying plant responses to environmental stressors [39]. Additionally, nitrate ($NO_3^-$) and nitrite ($NO_2^-$) present in PAW contribute as essential nitrogen sources required for the synthesis of proteins and other macromolecules, in addition to their signaling function. Several studies suggest that PAW can act as an alternative to chemical biostimulants, particularly in early plant developmental stages, such as seed germination [40]. However, the effectiveness of PAW is influenced by multiple factors, including the plant species, the specific chemical composition of PAW, and the experimental conditions under which PAW is applied [41].

Characteristics of tap water (control) and the three PAW types are shown in Table 1. A little reduction in pH was observed between the control and TS PAW, as well as FDBD; however, for MW PAW, pH was as low as 3.9. The little pH variation before and after plasma treatment may be attributed to the inherent hydrogen carbonate and bicarbonate buffer system in tap water, in contrast to the significant acidification reported during plasma treatments of deionized or distilled water [42], [43]. Our prior papers also confirmed these minor pH variations using tap water [18], [44], [45]. Overall pH changes were minimal in TS and FDBD, rendering this parameter a non-disruptive element in the germination process.

The EC and TDS increased after activation by plasma for all PAWs which is due to the formation and transport of active species and ions from plasma to PAW. Many previous works confirm this effect [46], [47]. Oxidation-Reduction potential (ORP) is an indicator that illustrates the propensity of a solution to become an oxidizing or reducing agent. As shown in Table 1, ORP of PAW increases after plasma treatment. Although TS-PAW exhibited the highest $H_2O_2$ concentration, its ORP remained relatively low. This is likely due to the near-neutral pH (~7.4), which reduces the oxidative contribution of $H_2O_2$, as its redox potential is pH-dependent. Additionally, the presence of reducing species such as $NO_2^-$ may also offset the oxidative strength. In contrast, MW-PAW showed a significantly higher ORP despite lower $H_2O_2$ levels, mainly due to its strong acidification and high $NO_3^-$ concentration. This highlights that ORP reflects the combined effect of multiple redox-active species and the solution pH. Although a slight temperature increase was shown right after plasma treatment, all PAWs were isothermal with the ambient environment before irrigation. The discharge power was adjusted to 52.5 W (21 × 2.5 W) for the TS plasma, 48.8 W for the FDBD plasma, and 1800 W for the MW plasma, with different volumes of water used in each case according to the plasma system configuration.

Table 1. Physio-chemical properties of PAW prepared using three plasma systems: transient spark (TS), fountain dielectric barrier discharge (FDBD), and microwave (MW). Tap water (TW) served as the untreated control. Values represent mean ± SD (n = 6) from independent preparations. No statistical comparison was applied due to the use of different plasma systems.

|      | pH | EC ($\mu S/cm$) | ORP (*mV*) | TDS (*ppm*) | Temp (°C) | Power (W) |
|------|-----|-----|-----|-----|-----|-----|
| TW   | 7.7±0.2 | 538.5±8.5 | 250±24 | 290.5±3.5 | 21.3±1.3 | - |
| TS   | 7.41±0.30 | 582.5±3.5 | 286.7±17.3 | 334.5±2.5 | 26.3±0.75 | 52.5 |
| FDBD | 7.35±0.25 | 569±3 | 261.05±37 | 312±11 | 24.0±0.25 | ٤٨,٨ |
| MW   | 3.95±0.05 | 747.5±10.5 | 402.9±23.2 | 409±4.9 | 23.0±0.3 | 1800 |

Note: ORP – oxidation-reduction potential; EC – electrical conductivity; TDS – total dissolved solids; Temp- Temperature

Figure 4 illustrates the concentrations of $H_2O_2$, $O_3$, $NO_2^-$, and $NO_3^-$ produced in the three types of PAW. The concentrations of $H_2O_2$ obtained were around 479 $\mu M$, 51 $\mu M$, and 3 $\mu M$ for TS, FDBD, and MW, respectively. The $O_3$ values for TS, FDBD, and MW PAW were only 6.7 $\mu M$,

2.7 $\mu M$, and 0.1 $\mu M$, respectively. The $NO_2^-$ values for TS, FDBD, and MW were around 1.32 $mM$, 0.26 $mM$, and 0.11 $mM$, respectively. The $NO_3^-$ values for TS, FDBD, and MW were around 1.94 $mM$, 0.93 $mM$, and 6.91 $mM$, respectively. All plasma sources are abundant sources of RONS transported into water, as previously shown [18], [44], [45], [48]. The almost stable pH (in TS and FDBD) ensures that these RONS remain intact for many hours post-plasma treatment, hence enhancing their applicability to plants, except for MW where pH was reduced to 3.9. The TS and FDBD PAW are similar in composition, but TS PAW shows higher concentrations of all species, while MW makes dominantly an acidic solution of $NO_3^-$.

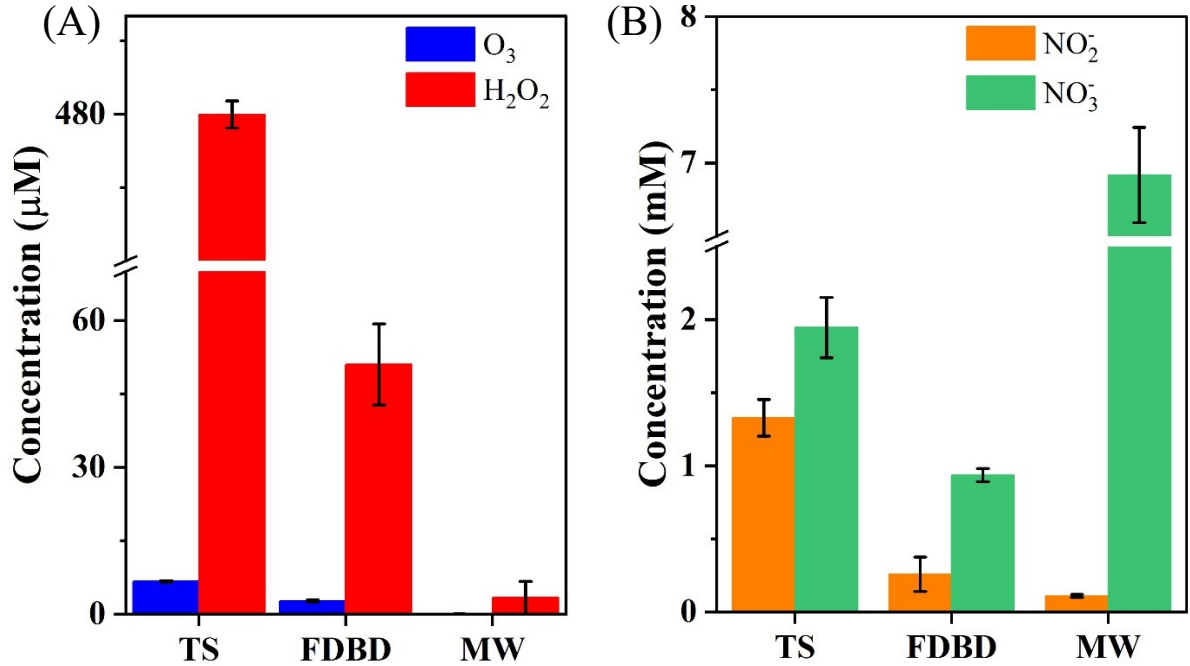

Figure 4. Concentrations of $H_2O_2$, $O_3$ (A), $NO_2^-$, and $NO_3^-$ (B), in plasma-activated water produced by TS, FDBD, and MW plasma. Values are presented as a mean ± standard deviation (n=6).

### 3.2. Seed germination in Petri dishes

Figure 5 A and B show the variations in Germination Percentages (GP) and Mean Germination Time (MGT) caused by various PAW. Although GP and MGT for MW PAW is higher, there is no significant difference between data. Figure 5 C illustrates the impact of PAW on seed vigor. The PAW has a significant effect on the vigor of the seeds. All groups showed significant differences compared to Tap water.

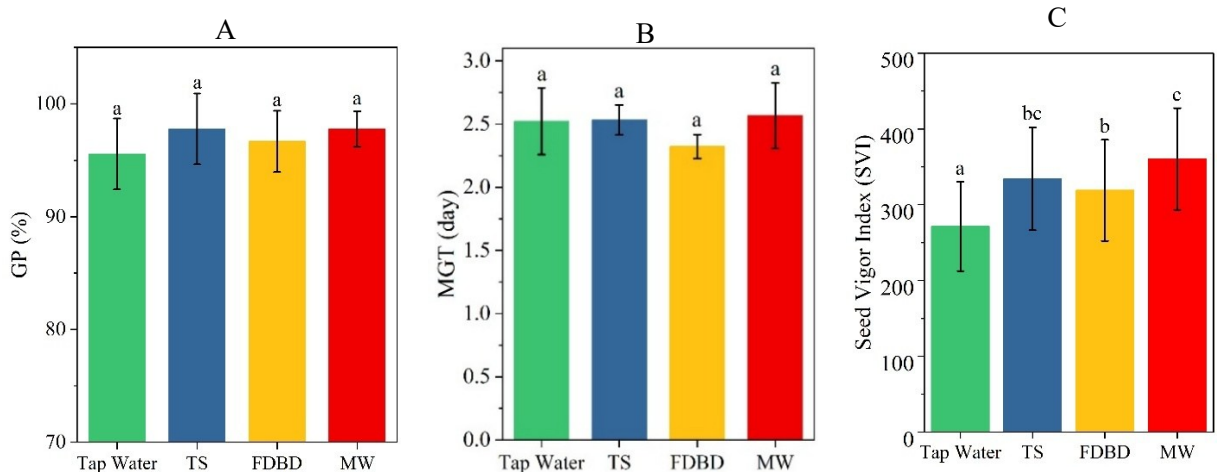

Figure 5. Seed germination parameters (A) GP, (B) MGT and (C) SVI. Values are presented as a mean ± standard deviation (n=3). Different lowercase letters indicate significant differences among the treatments at p < 0.05.

Dobrin et al.[49] reported that ROS induced by PAW may lead to seed coat cracking and thinning, thereby enhancing water and nutrient uptake. This process has been associated with improved germination parameters, including a higher germination rate, germination index, and vigor index. However, in our study, different PAW treatments did not significantly affect the percentage of germinated seeds or the mean germination time. Notably, significant changes were observed in the seed vigor index which reflects seed quality and energetic potential during germination and early growth stages.

Figure 6 illustrates the impact of PAW on the shoot and root lengths of *Lactuca sativa* L. seedlings on day 8. The shoot and root exhibited different responses to PAW treatment. A visual comparison of root morphology reveals that the root length was reduced in PAW-treated seedlings compared to those irrigated with tap water (hypocotyl necks are shown in Figure 6). However, the root length was not a determining factor for the growth and quality of leafy plants. During germination, seeds rely on their internal nutrient stores to support the embryo growth, which may explain the limited PAW effect on root elongation.

In contrast, shoot length showed a significant increase across all three PAW treatments. Seedlings treated with TS, FDBD, and MW PAW exhibited shoot length increases of 14.3%, 21.7%, and 22.5%, respectively, compared to the control.

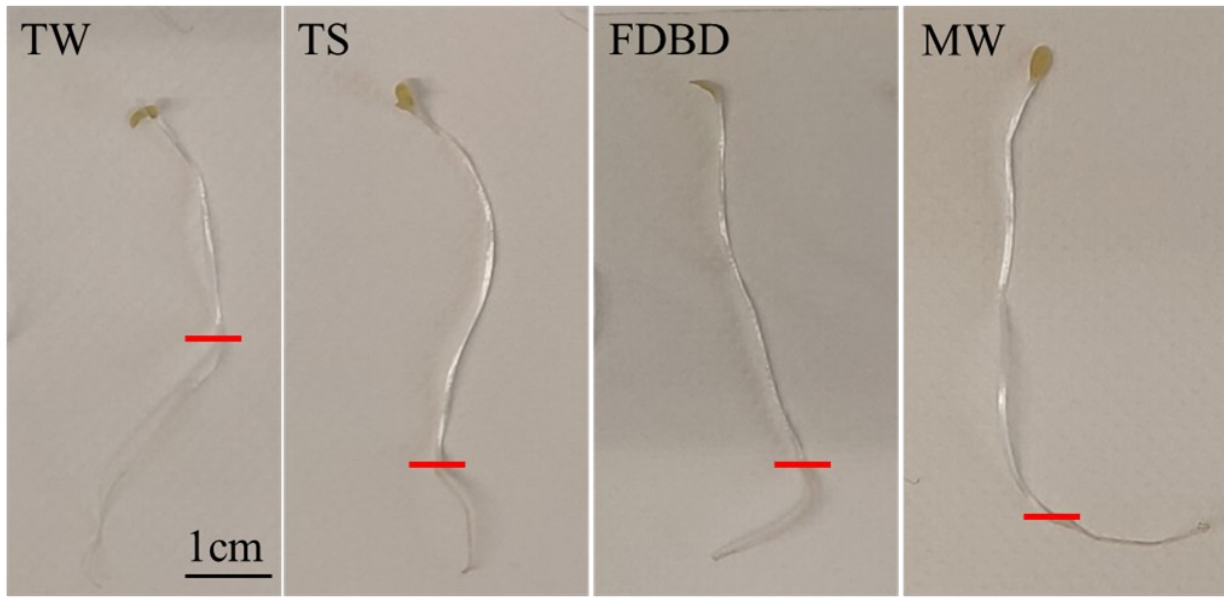

Figure 6. Photographs of seedlings germinated eight days after being sown. Red lines show transitions between the root and the hypocotyl. Scale bar 1cm.

Figure 7a compares the average shoot and root length of plant and a level of difference between data. Figure 7b shows the number of seedlings longer than 3 cm (upper part), which was shown to be most impacted by MW PAW, followed by TS, and then FDBD PAW. During the germination stage, seedlings need additional external nutrition for optimal plant growth. In PAW, $NO_3^-$ functions both as a signaling molecule and a chemical that may indirectly provide nutrients to seedlings. Consequently, the seedlings in all PAW samples exhibit superior growth indices relative to the tap water.

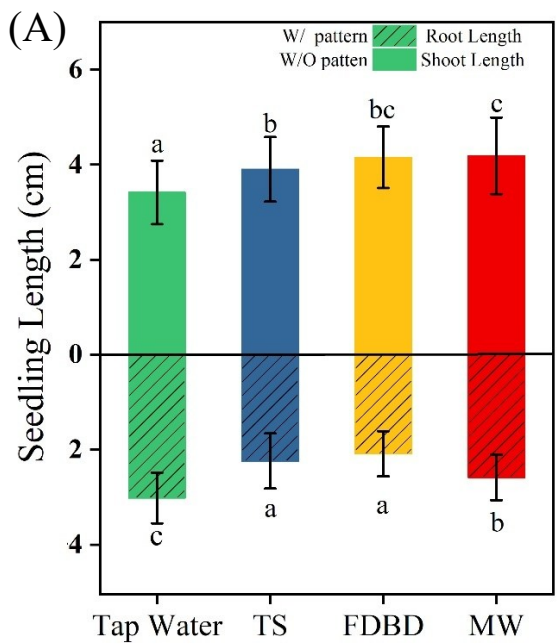
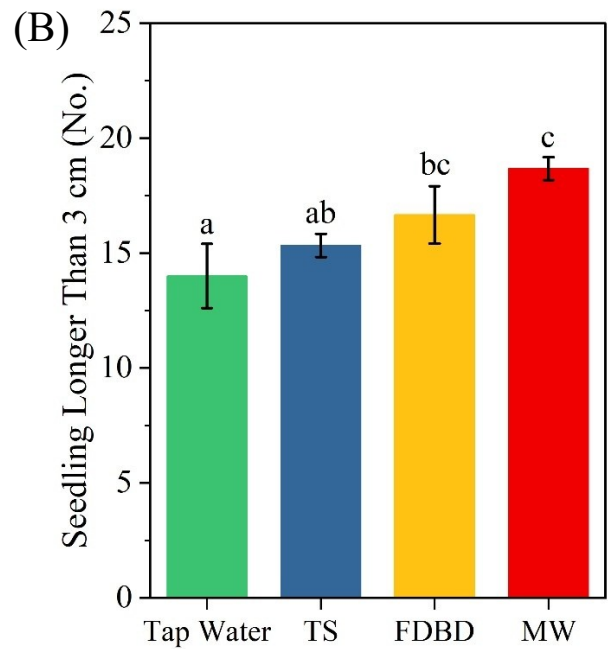

Figure 7. (A) Seedling length on 8th day after sowing. (B) number of seedlings longer than 3 cm. Values are presented as a mean ± standard deviation (n=3). Different lowercase letters indicate significant differences among the treatments at p < 0.05.

The germinating data were well adapted to the Richards function (Figure 8). The Richards function indicates that MW and FDBD exhibited a smoother initial phase of germination. All treatments reached their maximum germination rate on the 2nd and 3rd days. Population parameters $V_i$ (viability), $M_e$ (median germination time), $Q_u$ (dispersion) and SK (skewness) of the Richards' function are shown in Table 2.

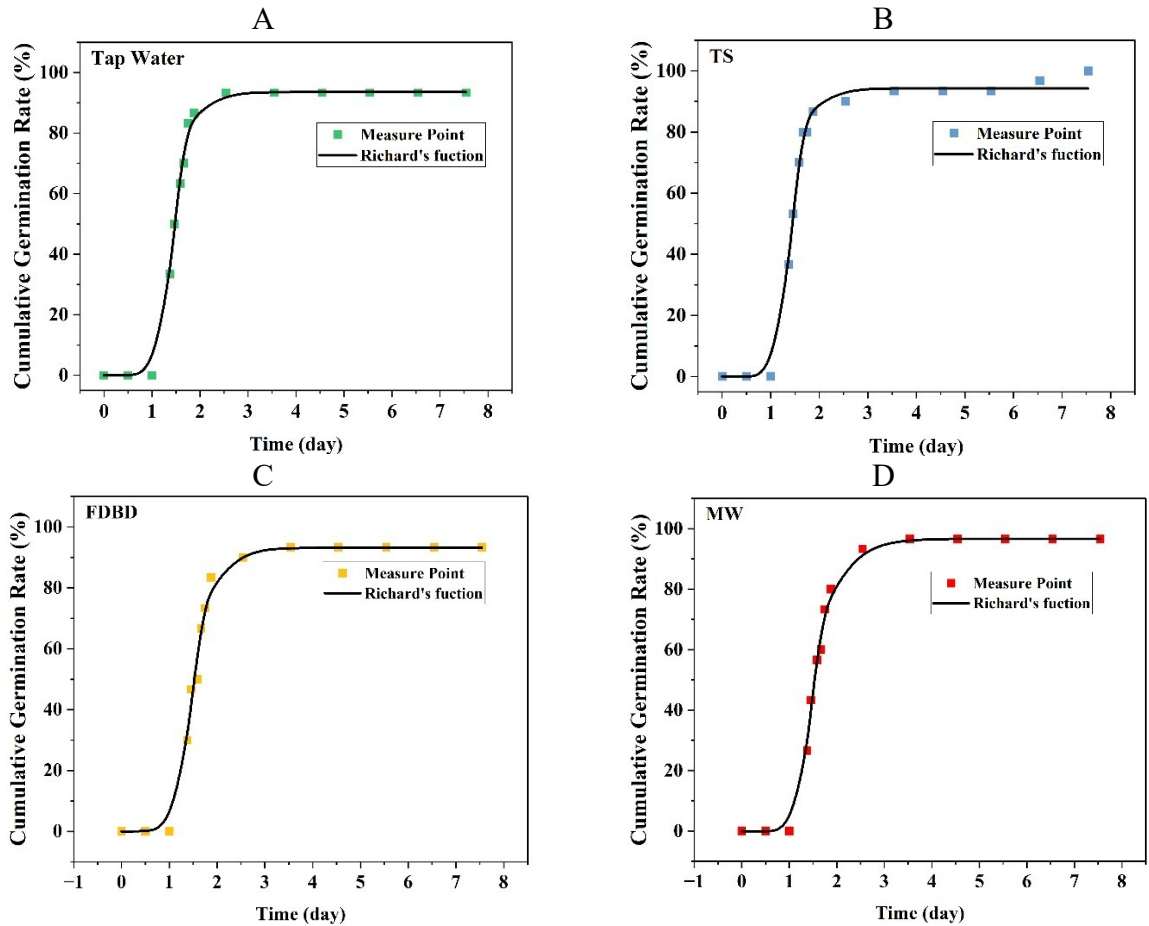

Figure 8. Richard's function fitted to the germination of lettuce seeds treated by (A) tap water, (B) TS, (C) FDBD, (D) MW

Analysis of the Richards function parameters showed no statistically significant differences in final viability among treatments. When analyzing the germination results (Table 2), the FDBD treatment showed the highest seed viability (97.06%) and the shortest median germination time (Me = 1.41 days), suggesting improved germination dynamics compared to other treatments. However, these differences were not statistically significant (p > 0.05), as indicated by shared superscript letters.

Table 2. The Richards' function for germination's population characteristics: Vi (viability), Me (median germination time), Qu (dispersion), and SK (skewness). Values are presented as a mean ± standard deviation (n=٣).

| Treatment group | Vi (%) | Qu (day) | Me (day) | Sk | $R^2$ Value |
|---|---|---|---|---|---|
| TW | 92.29±2.87[a] | 0.184±0.037[a] | 1.49±0.032[a] | 0.484±0.49[a] | 0.99 |
| FDBD | 96.04±3.38[a] | 0.199±0.027[a] | 1.41±0.0775[a] | 0.456±0.03[a] | 0.98 |
| TS | 96.044±5.38[a] | 0.152±0.0173[a] | 1.46±0.026[a] | 0.576±0.321[a] | 0.95 |
| MW | 95.56±3.51 [a] | 0.187±0.016[a] | 1.46±0.079[a] | 0.46±0.349[a] | 0.99 |

Among the treatments, seeds irrigated with MW PAW exhibited germination parameters comparable to those of the other plasma treatments, with a trend toward a smoother initial germination phase. These responses were associated with PAW containing the highest initial concentration of $NO_3^-$ and minimal concentrations of $O_3$ and $H_2O_2$. In contrast, TS and FDBD produced markedly higher levels of $H_2O_2$ and moderate amounts of $O_3$, yet germination parameters remained statistically similar to the control. This suggests that while plasma treatments did not significantly alter germination percentage or speed, the chemical composition of PAW may contribute to subtle differences in seedling vigor. Similar positive effects of PAW have been reported in previous studies. For instance, in 2011, Neumanová et al.[50] observed a 1.5-fold increase in coleoptile length in rye seedlings treated with PAW compared to untreated controls. Furthermore, Kučerová et al. [40] found that during the early growth stages, seeds primarily interact with $H_2O_2$ during imbibition and germination, whereas $NO_3^-$ and $NO_2^-$ are metabolized later, following germination initiation. The relatively low concentration of $H_2O_2$ in MW PAW may therefore have been beneficial, as $H_2O_2$ at low levels acts as a signaling molecule activating antioxidant defenses, while excessive levels—as in TS PAW—could become cytotoxic and hinder growth [51].

### 3.3. Seed germination and plant growth in soil

Lettuce seeds were planted and germinated directly in the pots with soil and were watered with PAW during 6 weeks, every 2-3 days: before germination 5 mL in each pot, second and third weeks 20 mL per pot and in the fourth week to last week 30 mL of PAW per pot was used for irrigation. The PAW irrigated plants exhibited considerably greater leaf sizes in comparison to the control, as demonstrated in Figure 9.

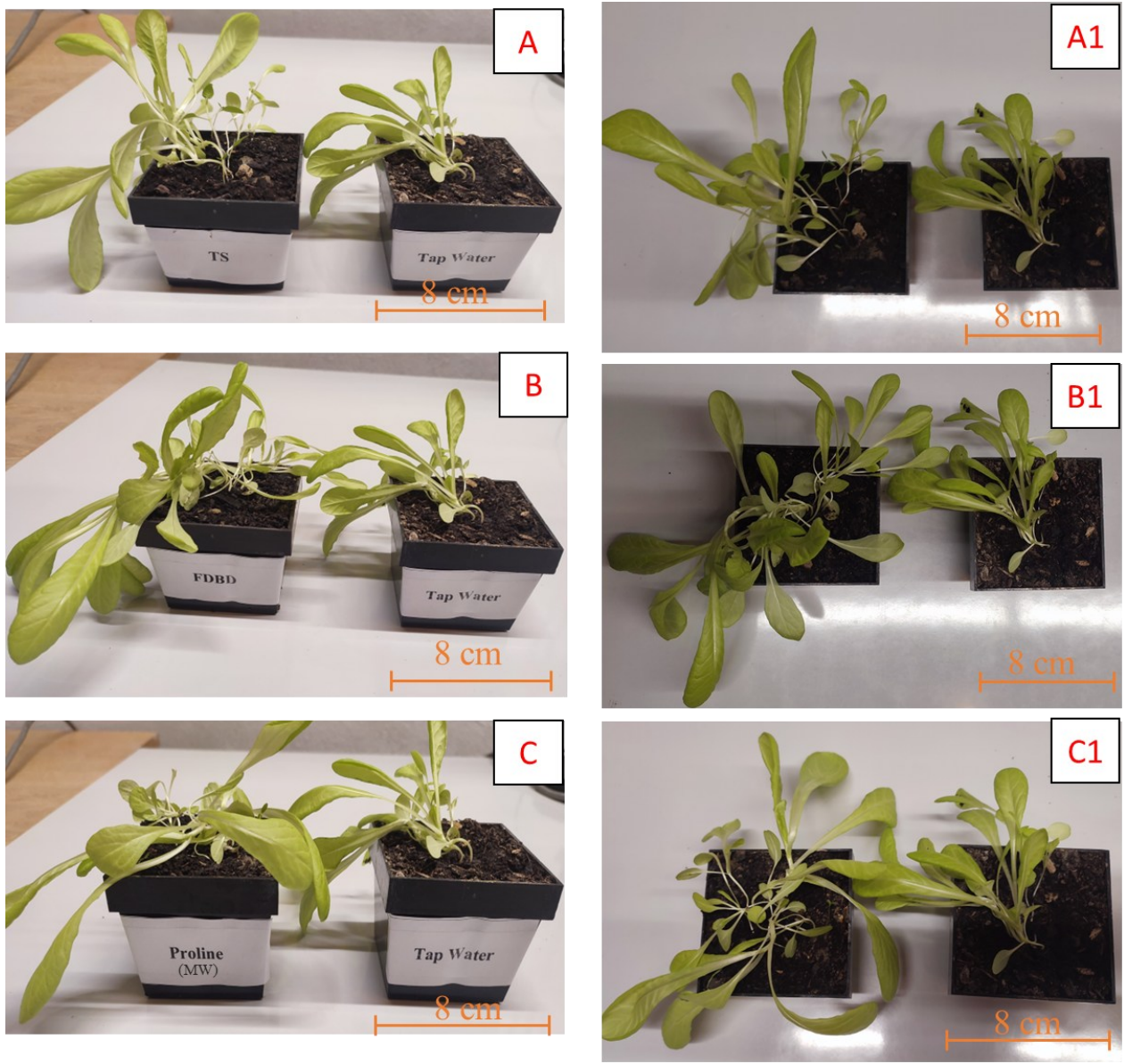

Figure 9. Photographs of lettuce plants after four weeks of cultivation in soil watered with tap water (control) and PAW. TS and tap water (A), FDBD and tap water (B), MW and tap water (C). A, B, C front view, A1, B1, C1 top-down view. Scale bar 8 cm.

The morphological characteristics of the plants, including the leaf dimensions, quantity, and developmental stage, were evaluated to assess the effects of PAW on the plant growth. The results indicate that lettuce plants irrigated with PAW exhibited considerably greater biomass accumulation compared to the control tap water irrigation. This enhancement in growth parameters suggests that PAW contributes to improved nutrient availability and physiological responses in plants.

As shown in Figure 10 A there is no significant difference in germination rate of seeds within 7 days after sowing into soil. While there is no difference in the quantity of leaves (for each pot) for

MW-activated water (2%), we observed a 20% increase in the leaf quantity in TS and FDBD groups which is shown in Figure 10 B.

There is an increase in the plant length of approximately 25% for FDBD and slightly above 30% equally for TS and MW compared with control (Figure 10 C). The dry weight more accurately represents the changes in biomass. The fresh and dry weights of the above-ground parts and roots of lettuce exhibited similar tendencies. Figure 10 D and E show 1.1, 1.15, 1.31-fold increase in the fresh weight, and 1.28, 1.35, 1.62-fold increase in the dry weight of plant when TS, FDBD and MW PAW were used for irrigation, respectively.

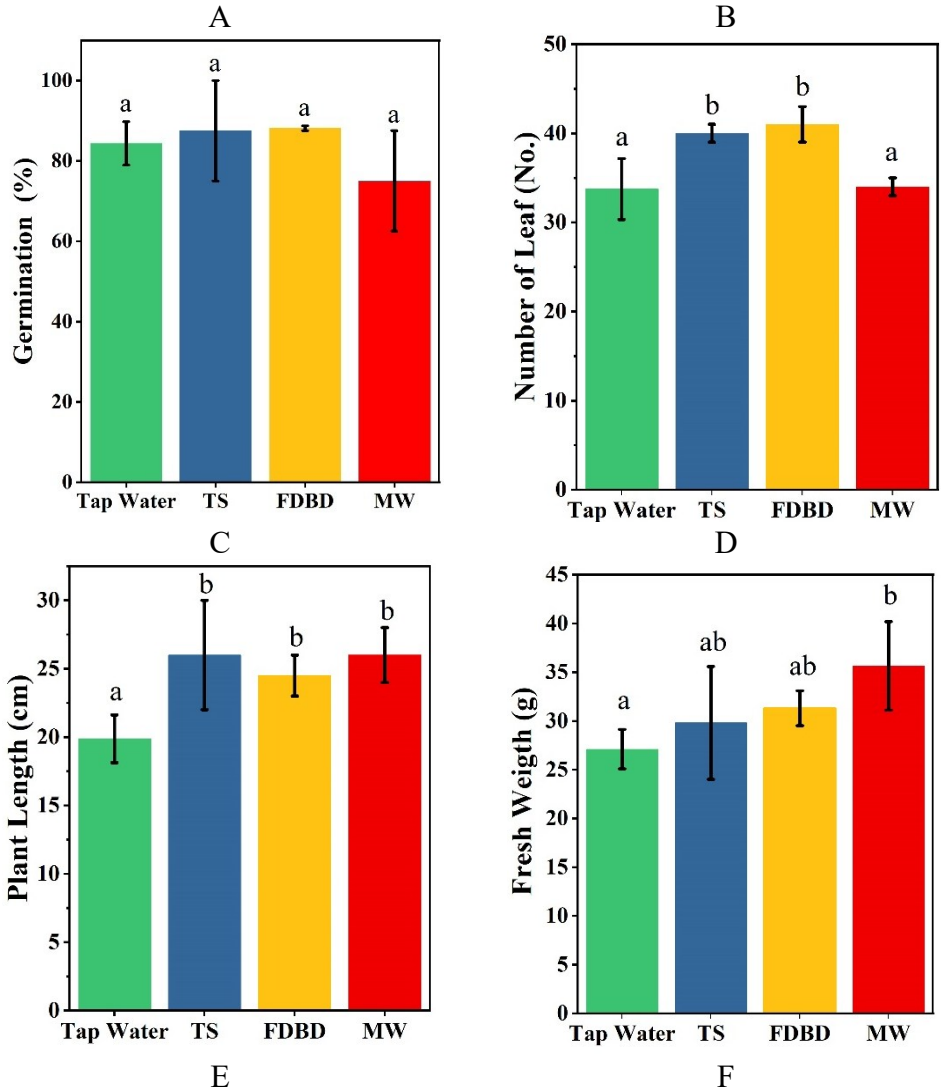

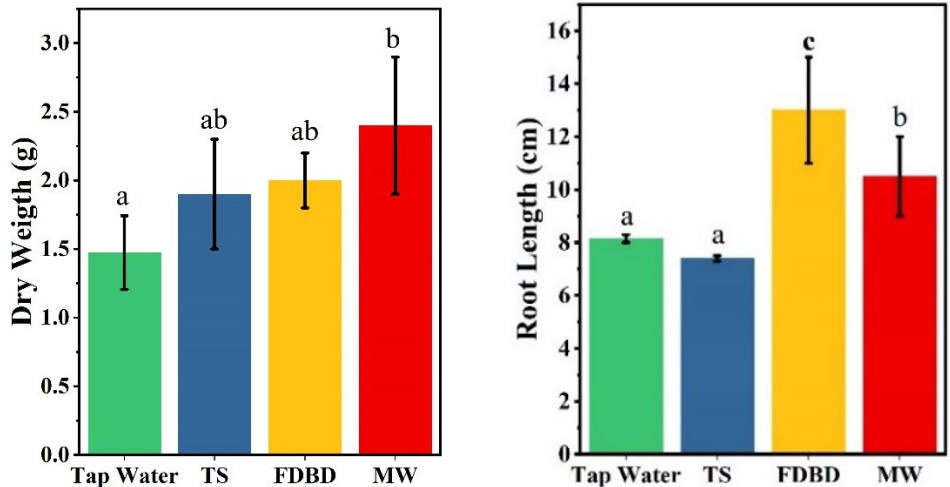

Figure 10. Plant growth parameters (A) Germination, (B) Number of leaves, (C) Plant length, (D) Fresh weight 6 weeks after sowing, (E) Dry weight 8 weeks after harvesting, (F) Root length. Values are shown as a mean ± standard deviation (n=12). Different lowercase letters indicate significant differences among the treatments at $P < 0.05$.

There is no significant difference in root length for TS with respect to control, but 1.59 and 1.28-fold increase showed by FDBD and MW. In addition, plant root volumes in MW and FDBD are higher in comparison TS and Control, as shown in Figure 10 F and Figure 11 specifically shows photographs of the roots of the representative plants in each experimental groups.

Nitrate ($NO_3^-$) is a key macronutrient, serving as the primary nitrogen source for plant growth while also functioning as a signaling molecule that regulates gene expression involved in nitrogen transport and assimilation [52]. Efficient nitrogen uptake is crucial for optimizing photosynthesis, metabolic activity, and overall biomass production. Our findings demonstrate a strong correlation between nitrate concentration and plant growth, supported by Pearson correlation coefficient of r=+0.982 for nitrate vs. fresh weight (FW) and r=+0.996 for nitrate vs. dry weight (DW). Among all treatments, only plants watered with MW PAW have shown a statistically significant enhancement in both fresh and dry weight, correlating with the highest $NO_3^-$ concentration recorded in this PAW. This suggests that MW-PAW irrigated plants not only benefited from enhanced nitrate availability but also exhibited greater water retention capacity, as indicated by the observed FW-DW ratio.

Nitrate availability, irrespective of its source—traditional nutrient solutions or PAW nitrate – significantly influences the key growth parameters, including plant height, root length, number of leaves, photosynthetic rate, and biomass production.

Interestingly, the TS and FDBD treatments were characterized by elevated $H_2O_2$ concentrations, which may have contributed to the improved growth. At low concentrations, $H_2O_2$ functions as a signaling molecule, activating antioxidant defense mechanisms that enhance the plant growth. Our results suggest that the $H_2O_2$ concentrations used in this study, ranging from sub micromolar levels in MW PAW to nearly 0.5 mM in TS PAW, had a stimulatory effect on the monitored growth parameters. Similar findings have been reported by previous studies in various plant species, including spinach, pepper, tomato, soybean, mung bean [53] [26], [54]–[56], as well as in the model plant *Arabidopsis thaliana* [57]. The positive effects of PAW treatment on biomass accumulation may be attributed to enhanced protein synthesis [41].

$NO_3^-$ serves as the primary nitrogen supply for plant synthesis of proteins and nucleic acids; hence, it is regarded as the principal element of the PAW that contributes to biomass enhancement. Furthermore, $H_2O_2$ may contribute to biomass increases by promoting lignification of plant tissues. $H_2O_2$ enhances the activities of key enzymes involved in the phenylpropanoid pathway, including phenylalanine ammonia-lyase (PAL), cinnamate 4-hydroxylase (C4H), and 4-coumarate-CoA ligase (4CL), leading to lignin accumulation. Additionally, $H_2O_2$ upregulates the activities of DNase, RNase, and caspase-3-like enzymes, which facilitate rapid lignification [58]. Moreover, $H_2O_2$ serves as a co-substrate in oxidation reactions during lignin polymerization, further reinforcing its role in structural reinforcement and biomass enhancement [59].

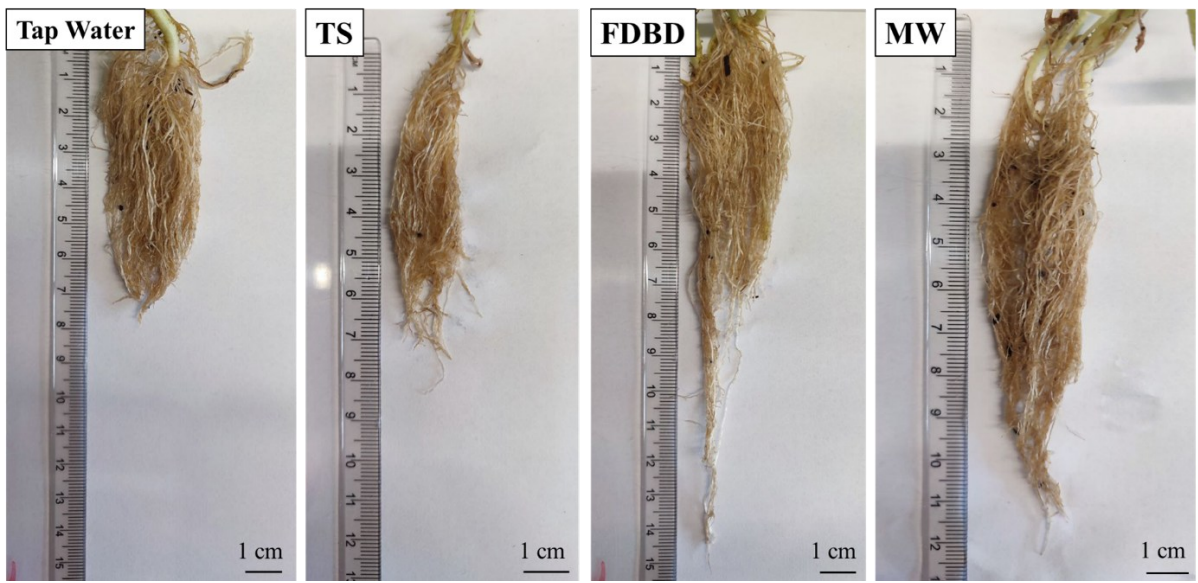

Figure 11. Root length of the harvested plant after 6 weeks in soil. Scale bar 1 cm.

The quantity of photosynthetic pigments, i.e., chlorophyll a, increased by 7% in plants irrigated by FDBD treated water as shown in Figure 12 A without pattern. Chlorophyll b was higher by 13% and 30% in plants irrigated with TS and FDBD PAW compared to control as shown in Figure 12 A with pattern. Even though these differences were statistically significant ($p < 0.05$), the magnitude of the change suggests that it had a small effect on pigment biosynthesis rather than a major impact. However, the MW treated water shows no significant difference in chlorophylls (Figure 12 A). TS and FDBD results show about 1.5-fold increase in carotenoids with respect to the control, while the MW treated water did not significantly increase carotenoids (Figure 12 B).

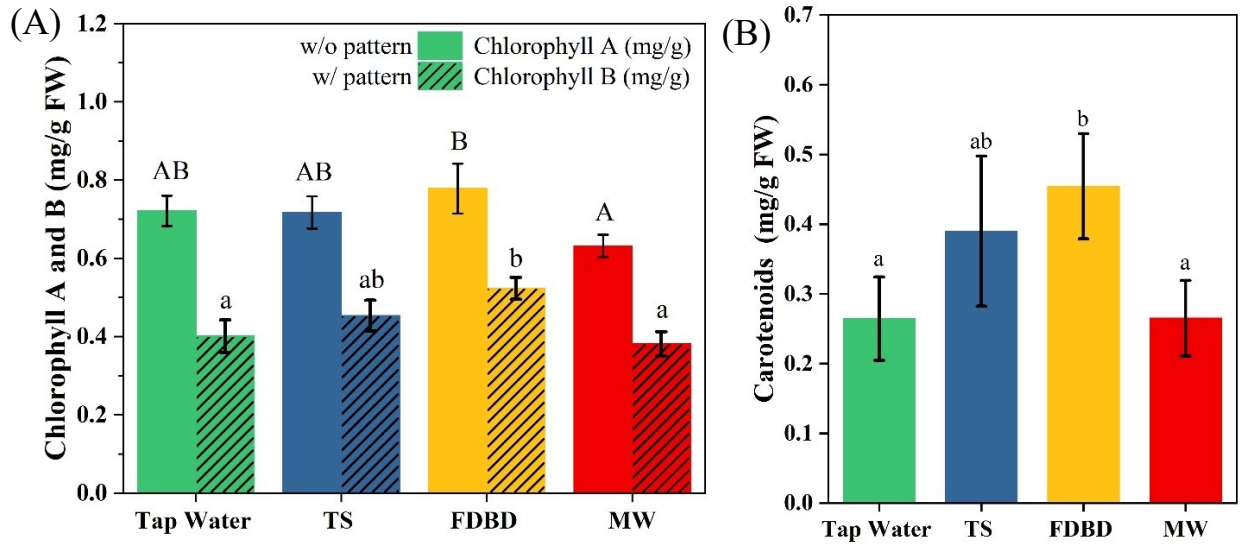

Figure 12. (A) Chlorophyll A and B of FW, without (w/o) and with (w/) pattern, respectively, (B) Carotenoids content in above-ground parts of lettuce plants after 6 weeks of soil cultivation irrigated with the PAW. Values are presented as a mean ± standard deviation (n=4). Different letters indicate significant differences among the treatments at $p < 0.05$.

Chlorophyll pigment content also indirectly provides information on mineral nutrition, as chlorophyll stores significant amounts of nitrogen. It is also a sign of overall plant metabolism [60], [61].

Investigating the effects of three different types of PAW on lettuce plants showed no significant impact on the production of photosynthetic pigment chlorophyll a. However, a significant difference in both chlorophyll b concentration and carotenoid synthesis was observed only under FDBD treatment. Carotenoids defend plants against photo-oxidation and have an additive function in photosynthesis. Their rise often signifies a rise in oxidative stress inside plant cells, which was higher in TS and FDBD PAW, since they had higher $H_2O_2$ concentrations. [62].

Although $H_2O_2$ and $O_3$ concentrations were the highest in the TS treatment, this variant also exhibited elevated levels of $NO_2^-$ and $NO_3^-$, which might serve as protective factors, mitigating the negative effects of $H_2O_2$. Conversely, in the FDBD treatment, $NO_2^-$ and $NO_3^-$ concentrations were lower. Nitrate can also enhance the production of specific enzymes, such as peroxidases, which help neutralize $H_2O_2$ by converting it into water [63]. The lower availability of these anions in the FDBD treatment might limit these protective mechanisms, causing the plant to rely more on carotenoids as antioxidants. This suggests that the different effects of PAW treatments on oxidative stress response may be mediated by the availability of nitrogen species, influencing the balance between enzymatic and non-enzymatic antioxidant systems. While the uptake of RONS (e.g., $NO_3^-$, $NO_2^-$, $H_2O_2$) by plant tissues was not directly quantified, the identified correlations between PAW composition and plant responses indicate a biological effect mediated by these reactive species. Changes in the pH and electrical conductivity of the water are important because they are caused by plasma-driven RONS formation (like the breakdown of nitric and nitrous acid). This means that they are an important part of the overall chemical profile of PAW. The synergistic effects of these species probably play a role in the stimulating effects seen in the plant growth.

Although we observed that certain PAW treatments with higher $NO_3^-$ or $H_2O_2$ concentrations were associated with improved germination and growth parameters, these effects should be interpreted cautiously. Since each plasma system generates a unique mixture of RONS, causal attribution to individual species is speculative. Further studies involving isolation experiments or synthetic PAW mimics (e.g., controlled solutions of $NO_3^-$ or $H_2O_2$) are needed to determine the specific contributions of individual reactive species to the observed plant responses. Further studies should also investigate the specific role of $NO_3^-$ and $NO_2^-$ in regulating peroxidase activity and how this interplay affects overall oxidative stress resilience in lettuce plants.

## 4. Conclusion

The effect of PAW produced by three different plasma sources on germination parameters, visual appearance, growth parameters, and physiological and biochemical characteristics of *Lactuca sativa* seedlings germinated in petri dishes, and of 6-week-old plants cultivated in soil, were examined. Each air plasma source (TS, FDBD and MW) produces different concentrations of RONS in PAW, which results in different effects on seedlings and plants. MW-PAW, which contained the highest nitrate levels but very low concentrations of $H_2O_2$ and ozone, resulted in the greatest improvements in fresh and dry biomass. This suggests that nitrate plays a key role as a nutrient and signaling molecule promoting biomass accumulation. In contrast, TS-PAW, with the highest $H_2O_2$ concentration and moderate nitrate levels, showed positive effects on shoot elongation and pigment accumulation, indicating that moderate levels of $H_2O_2$ may act as growth stimulants via redox signaling. However, our investigations indicate that PAWs have a negligible impact on lettuce seed germination parameters from petri dish germination.

These research results indicate that the impact of PAW on lettuce plants in soil surpasses that of germination and early seedling growth in petri dishes. Elevated amounts of RONS may be beneficial for some plant growth parameters; however, they might adversely affect others. This study findings indicate that PAW can favorably influence many physiological parameters of plants, especially increased nitrate promotes the plant growth and the total biomass. Nonetheless, the total impact of PAW on plant development in these experiments was rather limited, especially of TS and FDBD PAW that contained $H_2O_2$. Consequently, the optimization of the PAW composition, as well as observation of plant growth over its entire life cycle are necessary steps before its final upscaling for agricultural use.

### Acknowledgment

Funded by EU NextGenerationEU through the Recovery and Resilience Plan for Slovakia under the project No. 09I03-03-V03-00033 EnvAdwice and the Slovak Research and Development Agency APVV-22-0247.

### Conflict of Interest

The authors declare that they have no conflict of interest.